\documentclass[aps,pra,twocolumn,a4paper,showpacs,superscriptaddress,floatfix,10pt]{revtex4}
\usepackage[pdftex]{graphicx,color}
\usepackage{dcolumn}
\usepackage{bm}

\usepackage{amsfonts}
\usepackage{amssymb}
\usepackage{amsmath}
\usepackage{amsthm}

\usepackage{subfigure}
\usepackage{rotate}
\usepackage{dsfont}

\usepackage{pdfsync}

\let\oldmarginpar\marginpar
\renewcommand\marginpar[1]{\-\oldmarginpar[\raggedleft\tiny #1]%
{\raggedright\tiny #1}}

\def\KeyWord#1{$\backslash$\IfColor{$\!\!$\textRed{#1}\textBlack}{#1}$\!\!$}

\newcommand{\be}{\begin{equation} }
\newcommand{\ee}{\end{equation} }
\newcommand{\ba}{\begin{eqnarray} }
\newcommand{\ea}{\end{eqnarray} }
\newcommand{\n}{\nonumber \\ }

\newcommand{\bit}{\begin{itemize}}
\newcommand{\eit}{\end{itemize}}

\newcommand{\unit}{\mathds{1}}

\newcommand{\THH}{\mathcal{\tilde{H}}}

\graphicspath{{FinalFigures/}}

\begin{document}

\title{How does a locally constrained quantum system localize?}

\author{Chun Chen}
\affiliation{School of Physics and Astronomy, University of Minnesota, Minneapolis, Minnesota 55455, USA}

\author{Fiona Burnell}
\affiliation{School of Physics and Astronomy, University of Minnesota, Minneapolis, Minnesota 55455, USA}

\author{Anushya Chandran}
\email{anushyac@bu.edu}
\affiliation{Department of Physics, Boston University, Boston, MA 02215, USA}

\date{\today}

\begin{abstract}
At low energy, the dynamics of excitations of many physical systems are locally constrained. 
Examples include frustrated anti-ferromagnets, fractional quantum Hall fluids and Rydberg atoms in the blockaded regime.
Can such locally constrained systems be fully many-body localized (MBL)?
In this article, we answer this question affirmatively and elucidate the structure of the accompanying quasi-local integrals of motion.
By studying disordered spin chains subject to a projection constraint in the $z$-direction, we show that full MBL is stable at strong $z$-field disorder and identify a new mechanism of localization through resonance at strong transverse disorder.
However MBL is not guaranteed; the constraints can `frustrate' the tendency of the spins to align with the transverse fields and lead to full thermalization or criticality. 
We further provide evidence that the transition is discontinuous in local observables with large sample-to-sample variations. 
Our study has direct consequences for current quench experiments in Rydberg atomic chains. 

\end{abstract}

\maketitle

At low energy, the dynamics of many physical systems are restricted to Hilbert spaces with local constraints.
For example, the canonical spin-ice compound Dy$_2$Ti$_2$O$_7$ has Ising-like magnetic moments that obey a local ice rule at low temperature \cite{Castelnovo:2012lh, Gingras:2014aa}.
Electronic systems such as the fractional quantum Hall liquids and $p$-wave superconductors \cite{MooreRead,Ivanov,KitaevMajorana,OregMajorana,LutchynMajorana,Barkeshli:2014aa,Cheng:2012aa,Lindner:2012aa} are believed to host quasi-particles with non-Abelian statistics, which produce a topologically degenerate manifold of states within which the low energy dynamics are constrained.
Finally, Rydberg excitations of cold atomic chains \cite{Kaufman:2014aa,Bernien:2017aa} are energetically forbidden to occupy adjacent sites in the blockaded regime.

\begin{figure}[ht]
\centering
\includegraphics[width=0.95\columnwidth]{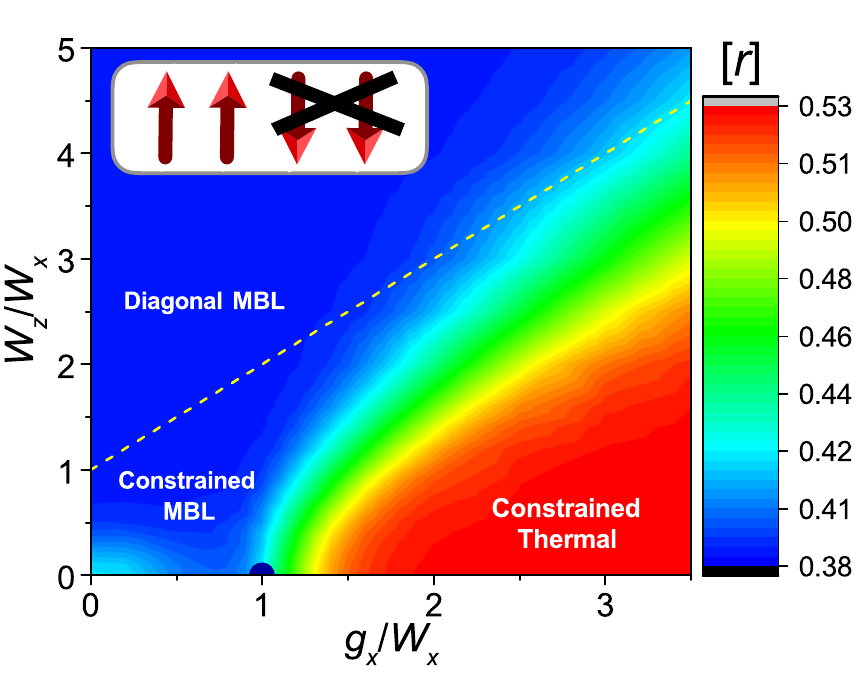} 
\caption{\label{rPhaseFig}  Infinite temperature dynamical phase diagram of the constrained Ising model in Eq.~\eqref{Ham} in which adjacent spins are forbidden to both point along $-z$ (inset). The mean level spacing ratio $[r]$ distinguishes the localized region with Poisson level statistics (blue, $[r]\approx 0.39$) from the thermalizing region with random matrix level statistics (red, $[r]\approx 0.53 $). 
At large $W_z/W_x$, the localization transition approaches the dashed line (see text), while at small $W_z/W_x$, the transition line intersects the $x$-axis at $g_x/W_x=1$ (black dot). }
\end{figure}

Little is known about the dynamical phases of locally constrained systems in isolation \cite{Chandran:2016ab,Lan:2017aa,Brenes:2017aa,Prakash:2017aa}.
Although their Hilbert space lacks a tensor product structure, there is a notion of locality because the influence of local measurements decay exponentially in space \cite{Chandran:2016ab}.
This suggests that constraints pose no impediment to local thermalization, as was numerically verified in pinned non-Abelian anyon chains \cite{Chandran:2016ab} and in dimer models \cite{Lan:2017aa}. 
But what of the effects of spatial disorder?
In unconstrained systems, quenched disorder can localize quantum particles and prevent the transport necessary for equilibration in isolation \cite{Anderson:1958ly}, a phenomenon known as many-body localization (MBL) \cite{Fleishman:1980hl,Altshuler:1997aa,Basko:2006aa,Gornyi:2005lq,Oganesyan:2007aa,Monthus:2010vn,Vosk:2013yg,Pekker:2014aa,Pal:2010gs,Znidaric:2008aa,Bardarson:2012kl,Serbyn:2013uq,Bauer:2013rz,Swingle:2013aa,Nandkishore:2014ys,Serbyn:2014aa,Iyer:2013aa,Kjall:2014aa,Laumann:2014aa,Luitz:2015fj,Chandran:2015aa,Tang:2015th,Michal:2014aa,Agarwal:2015aa,Vasseur:2015aa,Singh:2016aa,You:2016aa,Potter:2015ab,Nandkishore:2015aa,Altman:2015aa,Huse:2013kq,Serbyn:2013rt,Imbrie2016,Ros:2015rw,chandran2015constructing,Monthus:2016aa,Rademaker:2016aa,Nanduri:2014aa,Zhang:2015xy,Serbyn:2014ek,Yao:2015aa,Chandran:2014aa,Bahri:2015aa}.
In this article, we show that constraints pose no impediment to localization and present a model exhibiting new constraint-driven MBL and thermal phases (Fig.~\ref{rPhaseFig}).

Local constraints have two opposing effects on potentially MBL systems.
The constraints disallow certain intermediate states, blocking perturbative relaxation channels and
yielding more robust localization. 
This is the case for the `diagonal MBL' phase in Fig.~\ref{rPhaseFig}.  
However, when the constraints are  transverse to the disorder, they may frustrate localization by forbidding extremal eigenstates of the disorder potential, effectively decreasing the energy differences between adjacent spins.
This effect partly underlies the difficulty in localizing energy in pinned non-Abelian anyon chains \cite{Fidkowski:2009aa,Kraus:2011aa,Laumann:2012aa,Laumann:2012ac,Vasseur:2015ab,Potter:2016aa,Kang:2017aa,Prakash:2017aa}, and leads to the `constrained thermal' phase in Fig.~\ref{rPhaseFig}.
Surprisingly, such frustration does not preclude localization --  in the `constrained MBL' phase, the spins in regions with weak potential are pinned such that nearby spins may resonate and become approximate eigenstates of the transverse disorder potential without violating the constraints.
Thus, localization is favored by a mechanism reminiscent of `order-by-disorder' in frustrated spin systems \cite{Villain-J.:1980aa}.

Concretely, we study an open Ising chain of $N$ spins whose Hilbert space $\THH_N$ satisfies the constraint that neighboring spins cannot simultaneously point along $-z$ (see Fig.~\ref{rPhaseFig}, inset).
This Hilbert space describes quantum dimer ladders \cite{:aa}, pinned Fibonacci anyon chains \cite{Trebst:2008qf,Chandran:2016ab}, and Rydberg blockaded chains \cite{Pillet10}.
The dimension of $\THH_N$ is given by $F_{N+2}$, where $F_N$ is the $N$th Fibonacci number.
As $F_N \sim \phi^N$ with $\phi \equiv (\sqrt{5}+1)/2$ the golden mean for large $N$, the quantum dimension is irrational.
The Hamiltonian of the system is:
\begin{align} \label{Ham}
H &= \sum_{i=1}^N \left( g_i \tilde{X}_i + h_i \tilde{Z}_i \right )
\end{align}
where $g_i$ and $h_i$ are independently drawn on each site from box distributions $g_x + [-W_x, W_x]$ and $[-W_z, W_z]$ respectively, and $\tilde{X}_i = P \sigma_i^x P$, $\tilde{Z}_i = P \sigma_i^z P$ are the projected Pauli operators $\sigma^{x,z}_i$ on site $i$. The projection operator $P $ annihilates any $z$-configuration with the $\downarrow \downarrow$ motif,
\begin{align}
P = \prod_i \frac{(3 + \sigma_i^z + \sigma_{i+1}^z - \sigma_i^z \sigma_{i+1}^z)}{4},
\end{align}
so that $\tilde{X}_i$ can flip spin $i$ only if $\tilde{Z}_{i-1} = \tilde{Z}_{i+1} = 1$.
Without constraints, each spin independently precesses around its local field and there is neither transport of energy, nor local equilibration.  
The constraints force neighboring spins to interact as $[\tilde{X}_{i}, \tilde{X}_{i+1}] \neq 0$, producing the rich dynamical phase diagram in Fig.~\ref{rPhaseFig}.

\begin{figure}[ht]
\centering
\includegraphics[width=\columnwidth]{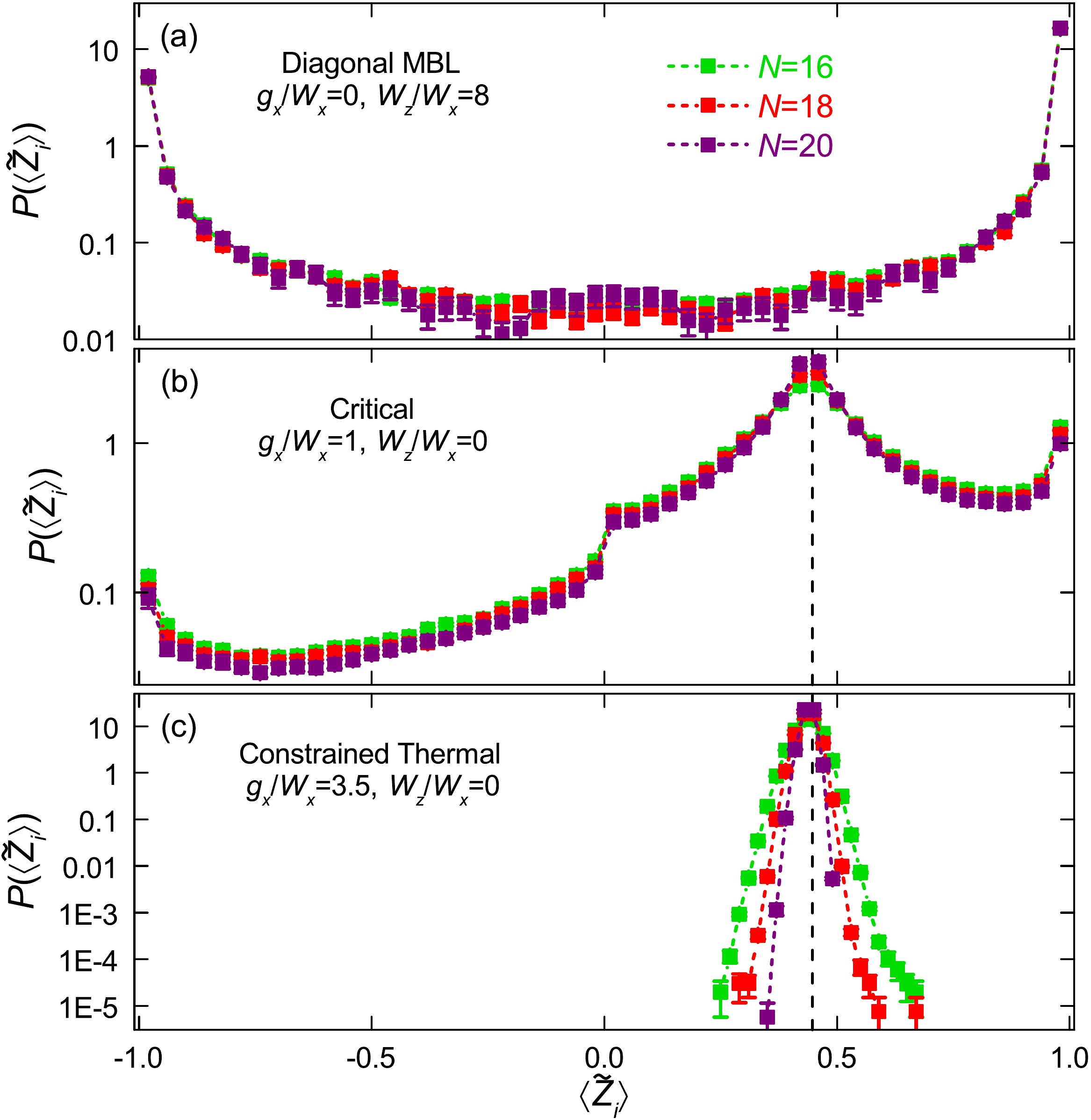}
\caption{\label{StrongZfig}   Probability distribution of $\langle E| \tilde{Z}_i |E \rangle$. In the diagonal MBL phase (panel (a)), the distribution is bi-modal with maxima near $\pm 1$. In the constrained thermal phase (panel (c)), the distribution concentrates around the infinite temperature average of $1/\sqrt{5}$ (dashed line). On the transition line in panel (b), the distribution shows features of the MBL and thermal phases. }
\end{figure}

{\it The diagonal MBL phase:} 
The restriction to $\THH_N$ is trivial when $H$ is diagonal in the $z$-basis ($g_i \equiv 0$).
This diagonal limit is trivially localized; every eigenstate $|E\rangle$ is uniquely labelled by the string of its $\pm1$ eigenvalues under the operators, $\tilde{Z}_i$ for $i=1, \ldots N$.  
The strings satisfy the constraint:
\begin{align}
|E\rangle = |\{\tilde{Z}_i \}\rangle, \quad\tilde{Z}_i \textrm{ and } \tilde{Z}_{i+1} \neq -1 \label{Eq:EigDiagProj}
\end{align}
The local conserved operators $\tilde{Z}_i$ define {\it $l$-bits}, which unlike their unconstrained counterparts, satisfy a restricted algebra wherein $\tilde{Z}_i \tilde{Z}_{i+1} = \tilde{Z}_i + \tilde{Z}_{i+1} - \unit$.
This implies \cite{:aa} that the $F_{N+2}$ tensor product operators which do not contain both $\tilde{Z}_i$ and $\tilde{Z}_{i+1}$ for any $i$ form basis for the space of conserved operators.

Imbrie \cite{Imbrie:2016aa,Imbrie2016} rigorously showed that the closely related unconstrained model:
\begin{align}
H_{IM} = \sum_{i=1}^N g_i \sigma_i^x + h_i \sigma_i^z + J_i \sigma_i^z \sigma_{i+1}^z.\label{Eq:HamImbrie}
\end{align}
 can be diagonalized using a sequence of quasi-local unitary operators $U$.  
 Since terms containing $r$ spins appear in the generator of $U$ with an amplitude that is (with high probability) exponentially small in $r$, the trivial $l$-bits at $g_i = 0$ extend to quasi-local $l$-bits at small $g_i$: $\tau_i^z \equiv U \sigma_i^z U^\dagger$.
These $l$-bits underlie the integrability, the dephasing dynamics and the low eigenstate entanglement of the fully MBL phase \cite{Huse:2013kq,Serbyn:2013rt,Huse:2014ac,Imbrie2016,Ros:2015rw,chandran2015constructing,Monthus:2016aa,Rademaker:2016aa,Pal:2010gs,Bauer:2013rz}.

The above arguments can be adapted to argue for full MBL in our model when $g_x, W_x \ll W_z$ (the upper-left corner of Fig.~\ref{rPhaseFig}).
Specifically, as (i) the constraints do not affect the level statistics of the spectrum of the diagonal limit at $g_i=0$, and (ii) certain terms only appear in the rotated Hamiltonian at higher order in perturbation theory as compared to the unconstrained case, we can perturbatively construct a quasi-local unitary $U$ which diagonalizes Eq.~\eqref{Ham} with high probability and defines $l$-bits:
$
\tilde{\tau}_i^z = U \tilde{Z}_i U^\dagger
$
with the same properties as Eq.~\eqref{Eq:EigDiagProj} in the diagonal limit.
$U$ also defines the quasi-local operator $\tilde{\tau}_i^x$ that flips the $z$-eigenvalue of the $l$-bit $i$: $\tilde{\tau}_i^x = U \tilde{X}_i U^\dagger$. As in the diagonal limit, $[\tilde{\tau}^x_i, \tilde{\tau}_{i\pm 1}^x] \neq 0$.

We support these claims with exact diagonalization performed on $N_s \geq 1000$ samples at $N=14-18$ and $N_s=500$ samples at $N=20$. 
Within each sample, we consider the central third of the sites and the central third of the eigenspectrum.
At large $W_z$, we expect that the $l$-bit $\tilde{\tau}_i^z$ is a weakly dressed version of $\tilde{Z}_i$ with a finite fraction of its operator weight on $\tilde{Z}_i$.
As $\langle E | \tilde{\tau}_i^z |E \rangle = \pm 1$, we expect a bi-modal distribution for $\langle E| \tilde{Z}_i | E \rangle$ with weight primarily at $\pm 1$ as $N \to \infty$.  
This is confirmed by Fig.~\ref{StrongZfig}(a).
The (small) weight between $-0.5$ and $0.5$ comes from eigenstates in which spins $i-1$ and $i+1$ point primarily along $+z$, so that spin $i$ points along/against its local field direction. As the local field is in the $x$-$z$ plane, the $z$-projection of spin $i$ is reduced \cite{:aa}.
 
\begin{figure}[ht]
\centering
\includegraphics[width=0.475\textwidth]{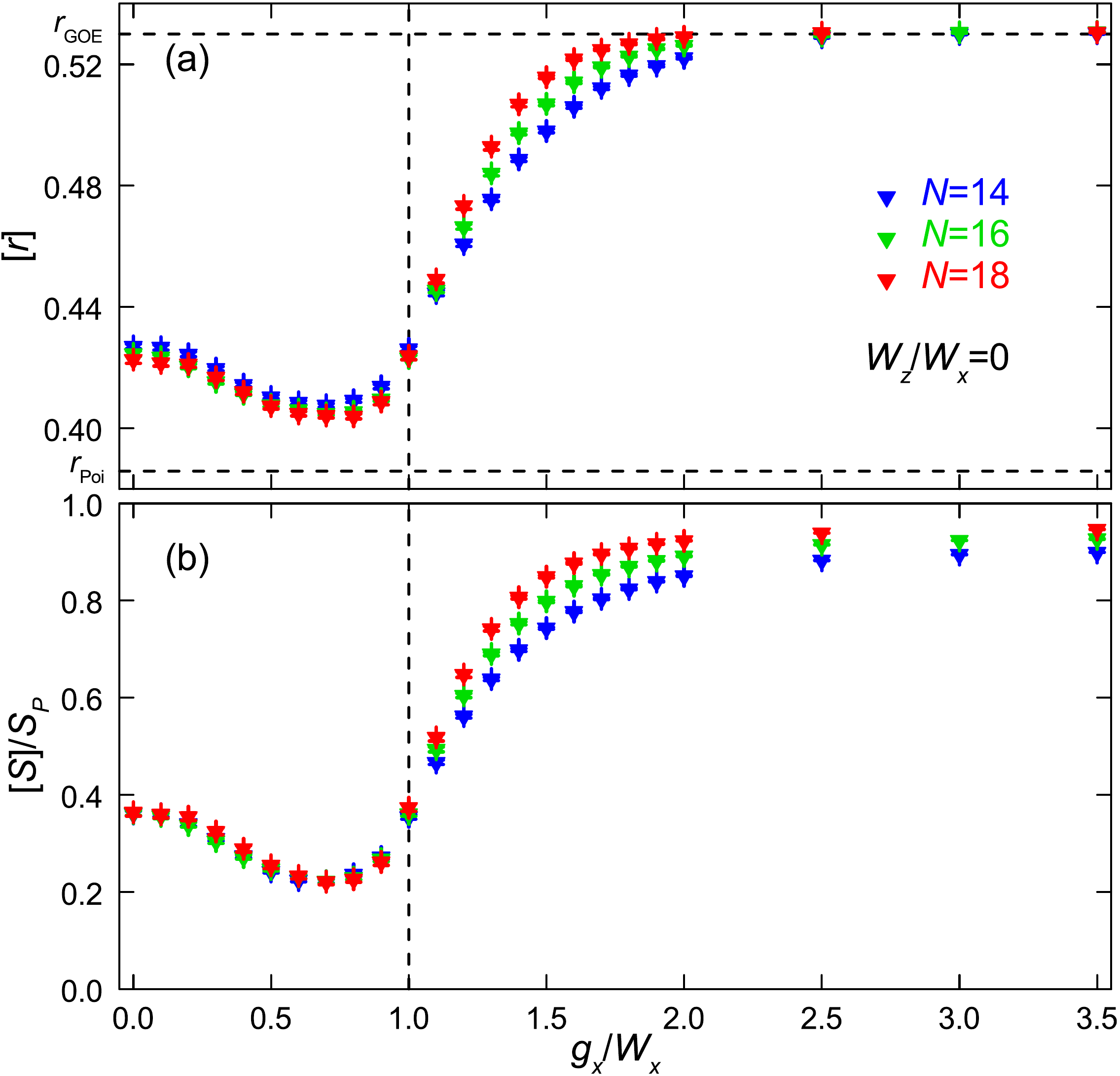}
\caption{\label{Fig3} (a) $[r]$ and (b) $[S]/S_P$ for $W_z = 0$. Both quantities
approach their thermal values with increasing $N$ for
$g_x/W_x > 1$, suggesting that the transition is at $g_x/W_x  = 1$ in the thermodynamic limit.  } 
\end{figure}

\emph{The constrained thermal phase:} 
When the maximum $x$-field $g_x + W_x$ is comparable to the typical $z$-field $\sim W_z$, the perturbative construction of the quasi-local unitary breaks down and a thermal phase emerges.
This suggests that the phase boundary lies at $W_z/W_x \sim g_x/W_x +1$ for large $W_z/W_x$ (dashed line in Fig.~\ref{rPhaseFig}).
The constrained thermal phase persists to small $W_z/W_x$ for $g_x/W_x \gg 1$, in agreement with the intuition that weakly disordered, strongly interacting models thermalize \footnote{There are however special points, like the translationally invariant model with $W_z=0$ and  $g_i\equiv g>0$, that may not be thermalizing. This may be the cause for the long-lived coherent oscillations in the $\uparrow \uparrow$ density observed in Ref.~\cite{Bernien:2017aa}.}.
Indeed Fig.~\ref{StrongZfig}(c) confirms that the model satisfies the eigenstate thermalization hypothesis (ETH) \cite{Jensen:1985aa,Deutsch:1991ss,Srednicki:1994dw,Rigol:2008bh,DAlessio:2016aa} expected of well-thermalized phases, in which individual eigenstates reproduce the expectation values of the thermal ensemble \cite{:aa}.
Specifically, panel (c) shows that the distribution of $\langle E | \tilde{Z}_i | E \rangle$ concentrates around the infinite temperature value of $1/\sqrt{5}$ with increasing $N$, in contrast to the distribution in the MBL phase in panel (a).

\emph{The constrained MBL phase:}
Finally, we turn to the regime in the lower left corner of Fig.~\ref{rPhaseFig} ($W_z/W_x, g_x/W_x \ll 1$) which exhibits numerical signatures of localization.
Here, random $x$-fields dominate on most sites, in contrast to the diagonal MBL phase where random $z$-fields dominate on most sites.
What is the mechanism underlying localization, given that adjacent spins {\it cannot} simultaneously align along the $x$-axis?

Consider a 3-site chain with $g_1\neq g_3>0$, but $g_2 = 0$.  
There are five eigenstates: four of them $|\tilde{X}_1 = \pm 1, \tilde{Z}_2=1, \tilde{X}_3=\pm 1 \rangle$ have energies $\pm g_1 \pm g_3$, while the fifth $|\tilde{Z}_1 = -1, \tilde{Z}_2=1, \tilde{Z}_3 = -1 \rangle$ has zero energy. 
In other words, the $x$ fields on sites $1$ and $3$ create an effective $z$ field of order $|g_3 - g_1|$ on site $2$.  
If $|g_2 |\ll |g_3 - g_1|$, $\tilde{Z}_2$ will be approximately conserved!
This suggests a resonance mechanism for localization in chains with interspersed strong and weak $x$-fields: the weak-field sites can be pinned in the $z$-direction to enable the strong-field sites to point along $\pm x$.

More formally, consider the `strong-strong-weak' (SSW) model obtained by repetitions of the 3-site $x$-field motif discussed above.
The SSW model is defined by Eq.~\eqref{Ham} with $h_i=0$ and
\be
g_i \in \begin{cases} [-\delta_w, \delta_w]  \ , \text{if } i =0 \text{ mod } 3 \\
[ 1 - \delta_s, 1+ \delta_s ] \ , \text{ otherwise. }
\end{cases}
\ee
For $\delta_w=0$, the local operators $\hat{O}_{3i} = g_{3i+1} X_{3i+1} + g_{3i+2} X_{3i+2}$ and $\tilde{Z}_{3i}$ commute with one another and with $H$.   
The eigenvalues $\epsilon_{3i} = 0, \pm g_{3i+1}, \pm g_{3i+2} ,\pm \sqrt{ g_{3i+1}^2 + g_{3i+2}^2 }$ of the $\hat{O}_{3i}$
 uniquely label the eigenstates, whose energies are given by $\sum_{3i} \epsilon_{3i}$.
Thus, this model is trivially localized with conserved operators given by $\{\hat{O}_{3i} \}$. 

In the supplementary material \cite{:aa}, we again adapt the methods of Ref.~\cite{Imbrie:2016aa,Imbrie2016} to argue that full MBL persists for  $\delta_w\ll \delta_s$ in the SSW model.
Specifically, as the energy spectrum obeys limited level attraction at $\delta_w=0$, and resonances can occur only if two nearby strong fields differ by an amount $\sim \delta_w$, there exists a quasi-local unitary that diagonalizes $H$ and defines a set of $l$-bits at small $\delta_w$.
Although the statistics of the random fields in Eq.~\eqref{Ham} differ from the SSW model, we expect the same mechanism to underlie localization when $g_x, W_z < W_x$, i.e. in the constrained MBL phase.
Note that this phase is smoothly connected to the diagonal MBL phase at $W_z/W_x \gg 1$.

\begin{figure}[tbp]
\centering
\includegraphics[width=\columnwidth]{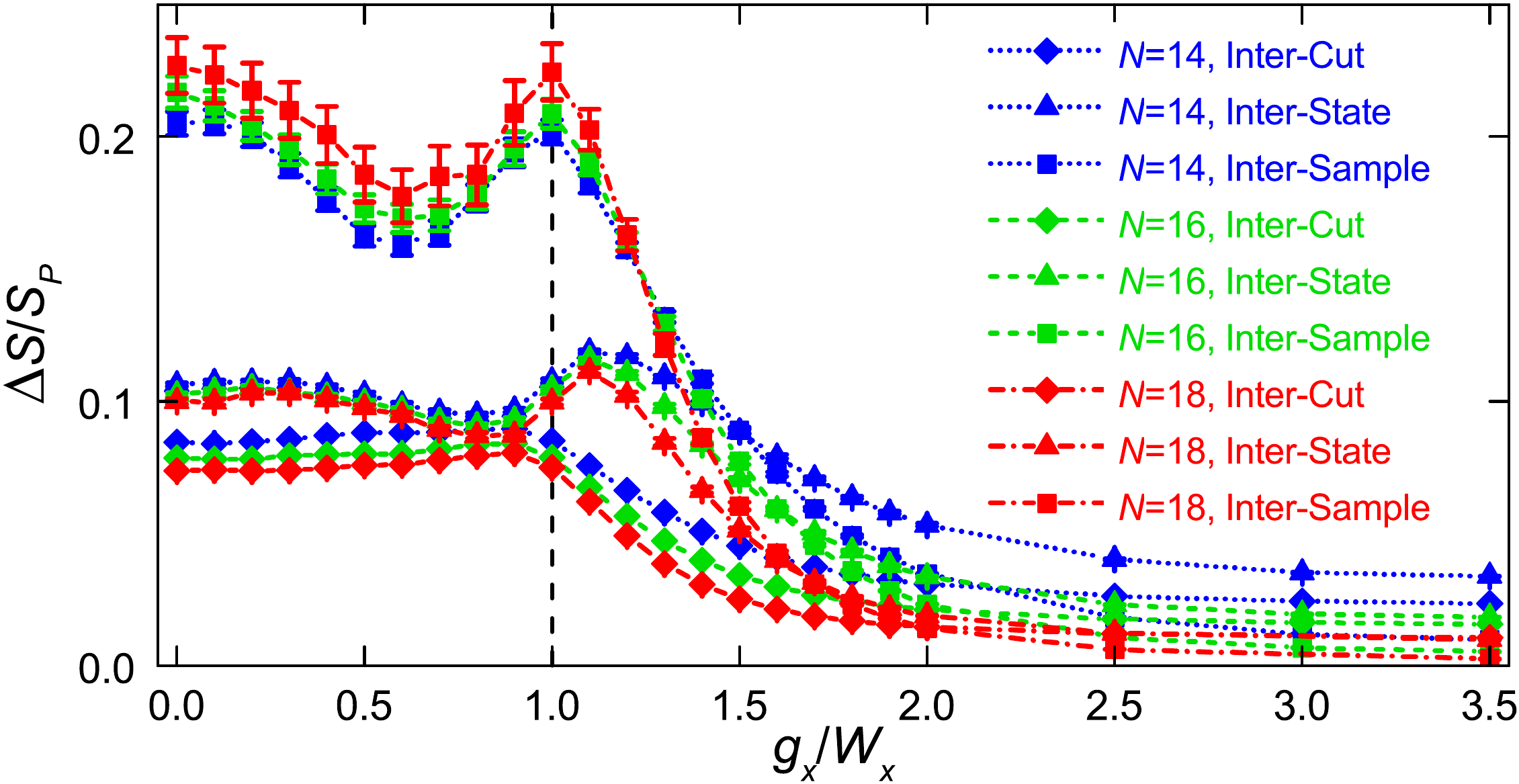}
\caption{\label{Fig4}  Normalized standard deviation of the half chain entanglement entropy $\Delta S/S_P$ at $W_z=0$ parsed by samples, states, and cuts, showing that the largest contribution comes from the inter-sample variance near $g_x/W_x=1$.}
\end{figure}

\emph{Behavior at $W_z=0$:} 
On the line $W_z=0$, the minimum $x$-field, given by $g_x-W_x$ controls the phase diagram: when $g_x>W_x$, there are no weak $x$-fields and we expect thermalization, while for $g_x<W_x$, we expect localization by the resonance mechanism.
We use two measures to test this: (a) the mean level spacing ratio $[r]$ \footnote{The level spacing $r$ is defined as $r(n) = \textrm{min}(\delta(n), \delta(n+1))/\textrm{max}(\delta(n), \delta(n+1))$ with $\delta(n) =E_n - E_{n-1}$ when the energies $E_i$ are enumerated in increasing order.} \cite{Oganesyan:2007aa,Atas:2013aa}, and (b) $[S]/S_P$, where $S$ is the half-chain entanglement entropy \footnote{The half-chain entanglement entropy is defined by $S \equiv -\textrm{Tr} \rho_L \log_2 \rho_L$, where $\rho_L = \textrm{Tr}_{\bar{L}} |E \rangle \langle E|$ is the reduced density matrix of the left half of the chain for eigenstate $|E\rangle$.}.
The mean is taken with respect to sites, states and samples and $S_P$ is the finite-size corrected entropy of an infinite temperature state \cite{Page:1993zf} \footnote{$S_P$ is the entanglement entropy of a random state in the Hilbert space with no constraint at the boundary between the left and right halves of the chain; for our model $S_P \equiv \log_2(\textrm{dim}(\rho_A)) -1/(2\ln 2) \equiv \log(F_{N/2+2}) - 1/(2\ln 2)$.   
Reinstating the constraint at the boundary numerically reduces the entropy by $\approx 0.06$, so that the entropy of a random state in $\THH$, or the mean entropy of infinite temperature eigenstates in eigenstate thermalizing phases, is $S_T \approx S_P - 0.06$.
As this correction is small, we use $S_P$ as a proxy for $S_T$ in the figure.}.
The finite-size flow of $[r]$ and $[S]/S_P$ in Fig.~\ref{Fig3} towards their respective thermal values of $r_{GOE} \approx 0.53$ and $1$ confirms the thermal phase for $g_x>W_x$.
In a localized phase, we expect $[r] \to r_{Poi} \approx 0.39$ and $[S]/S_P \to 0$ with increasing $N$.
While $[r]$ exhibits some finite-size flow towards $r_{Poi}$ for $g_x<W_x$, there is little flow in $[S]/S_P$.
We return to this issue below.  

At the purported transition $g_x / W_x =1$, the distribution of $\langle E| \tilde{Z}_i |E \rangle$ in Fig.~\ref{StrongZfig}(b) exhibit sharp features at four values: $\pm 1$, $0$ and at the infinite temperature value of $1/\sqrt{5}$. That is, spins either point along $\pm z$, $\pm x$, or are locally thermal in eigenstates. These features persist with increasing $N$, suggesting that (1) the transition point is heterogenous with respect to local observables, and (2) eigenstate expectation values of local observables jump across $g_x/W_x=1$. 

In order to determine the origin of the striking heterogeneity, we use different evaluations of the normalized standard deviation $\Delta S/S_P$, following Ref.~\cite{Khemani:2017aa}.
Fig.~\ref{Fig4} shows three possibilities: (1) inter-sample $\Delta_s [S]_{E,c}$, (2) intra-sample across eigenstates $[ \Delta_E [S]_{c}]_s$, and (3) intra-sample across the position of the cuts $[\Delta_c S]_{E,s}$, where $s, E, c$ respectively refer to sample, eigenstates, and cuts. 
In the thermal phase for $g_x/W_x>1$, all three quantities go to zero with increasing $N$, showing that $S$ for any position of the cut in any eigenstate in any sample is representative of the phase.
At $g_x/W_x=1$, on the other hand, the inter-sample deviation dominates and even slightly increases with $N$, while the other two variations decrease with $N$.
This strongly suggests that the transition is like an equilibrium first-order transition \cite{Fisher:1982aa,Pazmandi:1997aa} in which all of the variation in $S$ comes from the variation in the $x$-fields across samples.
This scenario has been suggested by a number of works \cite{Khemani:2017aa,Yu:2016aa,Dumitrescu:2017aa,Ponte:2017aa}, and our model provides the first clear microscopic observation.
We note that the increase in $\Delta_s [S]_{E,c}/S_P$ with $N$ cannot indefinitely continue as the variable is bounded. 

We return to the nature of the dynamical phase to the left of the transition.
Since performing local rotations about the $z$-axis on sites with $g_i <0$ yields an equivalent model with strictly positive $x$-fields,
 $g_x=0$ and $g_x=W_x$ represent the same dynamical transition: there is a sample at $g_x=W_x$ for every sample at $g_x=0$ with the same eigenstate properties.
The proximity to these transitions is likely the reason behind the lack of finite-size flow in Fig.~\ref{Fig3}.
As we argued for constraint-driven localization at small $W_z/W_x, g_x/W_x$, a natural conjecture is that the constrained MBL phase persists to $W_z=0$; we present analytic and numerical evidence in favor of this conjecture in forthcoming work \cite{Chen:aa}.

\emph{Discussion:}
We have studied the fate of MBL in a constrained disordered system, and found that constraints can either assist or frustrate localization, depending on whether or not they commute with the random fields. 
We have also shown by explicit construction that random fields in the transverse direction to the constraint lead to localization through a new resonance mechanism. 
Finally, we have provided strong evidence that the transition out of the thermal phase is `first-order', in the sense that the primary variation of the half-chain entanglement entropy comes from sample variation and that local observables vary discontinuously across the transition.

There are a number of interesting directions for further study.
First, the large sample-to-sample variations in $S$ at the numerically accessible system sizes suggest that the numerically extracted finite-size scaling exponents will obey the Harris criterion \cite{Chayes:1986kq,Chandran:2015aa}, in sharp contrast to the numerical exponents reported for unconstrained models \cite{Kjall:2014aa,Luitz:2015fj,Khemani:2017ab}.
More broadly, as the transition out of the thermal phase seems to be a consequence of weak fields (at least at $W_z=0$), the constrained Ising model provides a unique setting to isolate the effects of rare regions.
Quenches and other dynamical experiments should sharply bring out these rare region effects in energy transport and in entanglement growth.

Recently, Ref.~\cite{Potter:2016aa} argued that MBL is impossible in in pinned non-Abelian anyon chains, partly due to the lack of a tensor product Hilbert space.  
Since we have shown that constraints alone do not prevent localization, this delocalization must be a consequence of the $SU(2)_k$ symmetry respected by the anyon Hamiltonians. 
Constrained models may also provide simpler realizations of the non-ergodic delocalized phase posited in the non-Abelian systems \cite{Vasseur:2015ab}.

Finally, since the Hamiltonian generating the dynamics for recent quench experiments on atomic chains in the Rydberg blockaded regime \cite{Bernien:2017aa} has the form of Eq.~\eqref{Ham}, Fig.~\ref{rPhaseFig} is the dynamical phase diagram of disordered Rydberg chains. 
It would be extremely interesting to study the quench dynamics of the constrained thermal and MBL phases in these experiments. 

\emph{Acknowledgements:}
We are grateful to C.R. Laumann, S.L. Sondhi and D. Huse for stimulating discussions and for a careful reading of the manuscript, and to P. Crowley and V. Khemani for helpful conversations.
FJB acknowledges the financial support of NSF-DMR 1352271 and the Sloan Foundation FG-2015-65927.
\bibliography{paper-master}

\appendix

\section{A basis for the vector space of conserved operators from $l$-bits} \label{lbitsApp}

In unconstrained systems, the set of tensor products $Z_{i_1} ... {Z}_{i_k}$ is an algebraically complete basis for  all operators that commute with all $z$ fields.  In this appendix, we identify the subset set of these that comprise a complete basis for the operators that commute with $H = \sum_{i=1}^N h_i \tilde{Z}_i$. 

Consider the set $\mathcal{I}_N$ of tensor product operators $\tilde{Z}_{i_1} \otimes \cdots \otimes \tilde{Z}_{i_k}$, such that $1\leq  i_1<i_2  \ldots< i_k\leq N$, $0\leq k \leq (N+1)/2$ and $i_{a+1} \neq i_a +1$. Its cardinality is $F_{N+2}$. Here we show that the elements of $\mathcal{I}_N$ span the vector space of commuting operators on a length $N$ chain, and are linearly independent.

\emph{Spanning:} Any operator $\mathcal{O}$ that commutes with $H$ can be expressed as:
\begin{align}
\mathcal{O} = \sum_{m} c_{m} |E_m\rangle \langle E_m|
\label{Eq:ExpansionO}
\end{align}
where $c_m$ are arbitrary complex numbers and $|E_m\rangle $ are the eigenstates of $H$. For $H = \sum_i h_i \tilde{Z}_i$, these eigenstates are product states in the $z$-basis that satisfy the constraint that no two adjacent spins point along $-z$. Thus, the above relation can be re-written as:
\begin{align}
\mathcal{O} = \sum_m c_m \prod_{i=1}^N \left( \frac{1 + \gamma_i^m \tilde{Z}_i}{2} \right)
\end{align}
where $\gamma_i^m = \pm 1$ represent the $z$-projection of the $i$th spin in the $m$th eigenstate.
Multiplying out the products for each $m$, we find that any $\mathcal{O}$ can be expressed as a linear combination of the $2^N$ operators: $\tilde{Z}_{j_1} \otimes \ldots \otimes \tilde{Z}_{j_k}$ for $1\leq  j_1<j_2  \ldots< j_k\leq N$ and $0 \leq k \leq N$. 
For every product operator $\tilde{Z}_{j_1} \otimes \ldots \otimes \tilde{Z}_{j_k}$, we identify neighboring sites acted on by $\tilde{Z}$ and use the relation:
\begin{align}
\tilde{Z}_i \tilde{Z}_{i+1} = -\unit + \tilde{Z}_i + \tilde{Z}_{i+1},
\end{align}
repeatedly to re-write the operator as a linear combination of product operators in which $\tilde{Z}_i$ and $\tilde{Z}_{i+1}$ does not appear in the product for any $i$.
These operators are precisely the elements of the set $\mathcal{I}_N$.
Thus, any $\mathcal{O}$ that commutes with $H$ can be expressed as a linear combination of the operators in $\mathcal{I}_N$; that is, that the set $\mathcal{I}_N$ spans the vector space of commuting operators.

\emph{Linear independence:} 
Suppose the sets $\mathcal{I}_{N-1}$ and $\mathcal{I}_{N-2}$ are linearly independent.
If the only solution to the linear relation,
\begin{align}
 \left[\sum_{m=1}^{F_{N}} \alpha_{m} \mathcal{I}_{N-2}(m) \right] \otimes \tilde{Z}_{N}+ \sum_{n=1}^{F_{N+1}} \beta_{n} \mathcal{I}_{N-1}(n) = 0 
 \label{Eq:NLinDep}
\end{align}
has all real coefficients $\alpha_m, \beta_n$ equal to zero, then the set $\mathcal{I}_N$ is linearly independent.
Above, we have used the fact that every element of $\mathcal{I}_N$ is either an element of $\mathcal{I}_{N-1}$ or is the tensor product of an element of $\mathcal{I}_{N-2}$ with $\tilde{Z}_N$.

As the projectors onto the eigenstates ($z$-product states that satisfy the constraint) form a basis for the vector space of commuting operators at each $N$, we can express every element of $\mathcal{I}_N$ as a linear combination of these operators.
This allows us to replace elements of $\mathcal{I}_{N-2}$ and $\mathcal{I}_{N-1}$ with eigenstate projectors in Eq.~\eqref{Eq:NLinDep}: 
\begin{align}
 \left[\sum_{m=1}^{F_{N}}  C_{m} |E^{N-2}_m \rangle \langle E^{N-2}_m | \right]& \otimes \tilde{Z}_{N}+ \sum_{n=1}^{F_{N+1}}  D_{n} |E^{N-1}_n \rangle \langle E^{N-1}_n |= 0 
 \label{Eq:NProjExp}
\end{align}
where $|E^{N}_m \rangle$ is the $m$th eigenstate of the $N$ site chain and $C_m, D_n$ are related by an \emph{invertible} linear transformation to the $\alpha$ and $\beta$ coefficients:
\begin{align}
C_m &= \sum_{j=1}^{F_{N}} G^{N-2}_{mj} \alpha_j \nonumber \\
D_n &= \sum_{j=1}^{F_{N+1}} G^{N-1}_{nj} \beta_j
 \label{Eq:LinearTransCDalphabeta}
\end{align}
The invertibility of matrix $G^{N-2}$ (and analogously $G^{N-1}$) follows from the observation that the set of projectors onto the eigenstates $\{ |E^{N-2}_m \rangle \langle E^{N-2}_m|\}$ and the elements of $\mathcal{I}_{N-2}$ (by assumption) form bases for the vector space of commuting operators on $N-2$ sites. 

Now consider the action of the LHS of Eq.~\eqref{Eq:NProjExp} on states $|E_k^{N-2}\rangle |\uparrow, \uparrow\rangle$ and $|E_k^{N-2}\rangle |\uparrow, \downarrow\rangle$ for each $k=1, \ldots F_N$. This gives:
\begin{align}
 C_k + D_{k'} =0, \qquad
C_k - D_{k'} = 0 
\end{align}
where $k'$ labels the eigenstate $|E_{k'}^{N-1} \rangle = |E_k^{N-2} \rangle |\uparrow \rangle$ of the $N-1$ site chain. Thus, $C_k =D_{k'}=0 $ for all $k$. This reduces Eq.~\eqref{Eq:NProjExp} to:
\begin{align}
 \sum_{n=1}^{F_{N+1}} D_{n}  |E^{N-1}_n \rangle \langle E^{N-1}_n |= 0 
\end{align}
where for $n \in \{ k' \}$, $D_n =0$.
The above relation can only hold if each $D_n=0$ as the set $\{|E^{N-1}_n \rangle \langle E^{N-1}_n |\}$ is linearly independent.
Thus:
\begin{align*}
C_m = 0 ,\quad D_n = 0 \quad \textrm{for }1\leq m \leq F_{N}, \, 1\leq n \leq F_{N+1}
\end{align*}
Finally, inverting the linear transformation in Eq.~\eqref{Eq:LinearTransCDalphabeta}, we obtain:
\begin{align*}
\alpha_m = 0, \quad \beta_n = 0 \quad \textrm{for }1\leq m \leq F_{N}, \, 1\leq n \leq F_{N+1}
\end{align*}
Thus, the only solution to the linear relation in Eq.~\eqref{Eq:NLinDep} has zero coefficients and the set $\mathcal{I}_N$ is linearly independent.

It is easy to show that $\mathcal{I}_{1}$ and $\mathcal{I}_2$ are linearly independent. Thus, by induction, $\mathcal{I}_{N}$ is linearly independent for all $N$. 

\section{The Strong-Strong-Weak (SSW) model} \label{SSWApp}
Here we provide some additional details of the SSW model with random fields only in the $x$ direction which admits a localized phase. Recall that the Hamiltonian has the form
\begin{align} \label{SSW}
H &= \sum_{i} g_i \tilde{X}_i  \n
g_i &\in \begin{cases} [-\delta_w, \delta_w]  \ , \text{if } i =0 \text{ mod } 3 \\
[ 1 - \delta_s, 1+ \delta_s ] \ , \text{ otherwise }  
\end{cases}
\end{align}
We will take $\delta_s<0.17$ for simplicity. 

To understand the model's spectrum, we begin from the limit $\delta_w =0$.  
Since $[ \tilde{X}_{i+2}, \tilde{X}_i ] =0$, we need only to understand the possible eigenvalues $\epsilon_{3i}$ of 
\be 
\hat{O}_{3i} = g_{3i+1} \tilde{X}_{3i+1} + g_{3i+2} \tilde{X}_{3i+2} \ .
\ee
Consider spins $(0,1,2,3)$ with $g_0 = g_3 =0$, and $g_1\neq g_2 \neq 0$.  Because of the constraint, the possible eigenvalues of $\hat{O}_0$ depend on the values $(\tilde{Z}_0, \tilde{Z}_3)$.   
Let $\tilde{Z}_0$ be fixed by the boundary condition.  Then the  possibilities are:
\begin{align}
\tilde{Z}_0 = -1: & \begin{cases}  \tilde{Z}_3 = -1  & \ \ \epsilon_{0} = 0  \\
\tilde{ Z}_3= 1  & \ \ \epsilon_{0} = \pm g_2 
 \end{cases}
 \n
\tilde{Z}_0=1: & \begin{cases}  \tilde{Z}_3= -1  & \ \ \epsilon_{0} = \pm g_1 \\
\tilde{Z}_3 = 1 & \ \ \epsilon_{0} = 0 , \ \pm \sqrt{g_1^2 + g_2^2} 
\end{cases}
\end{align}

In either case, once we have specified the boundary condition and $\epsilon_{0}$, the value of $\tilde{Z}_3$ is uniquely determined.  
Next, when we consider $\epsilon_{3}$, whose possible values depend on the values of $\tilde{Z}_3$ and $\tilde{Z}_6$, we should treat $\tilde{Z}_3$ as fixed.  
Consequently, once we have fixed the boundary condition and $\epsilon_{0}$,  the possible values of $\epsilon_{3}$ are non-degenerate, and choosing one of these fixes the value of $\tilde{Z}_6$-- and so on for the remaining $\epsilon_{3i}$.  
In other words, provided that the non-zero couplings are all distinct, choosing $\{ \epsilon_{3i} \}$ uniquely fixes the eigenstate, pinning the value of $\tilde{Z}_{3i}$ for each $i$.   

Thus for $\delta_w =0$ with fixed $\tilde{Z}_0$, the SSW model has a non-degenerate spectrum with eigenvalues determined by the uncorrelated values of the strong random fields. This leads to Poisson level statistics.  In particular, it satisfies the criterion of limited level attraction that is required by Imbrie \cite{Imbrie:2016aa, Imbrie2016} in his proof that many-body localized phases exist.

From this starting point, we may turn on $\delta_w$ and assess the impact of the weak fields on the eigenstates perturbatively in $\delta_w/{\delta_s}$, following Refs.~\cite{Imbrie:2016aa, Imbrie2016}. Though we will not replicate the details of this construction for our model, we expect that it is very similar to that in the above references. The relevant similarities are:
  
(1) Probability of resonances at first order in perturbation theory: A first order resonance occurs if two spin configurations connected by $\tilde{X}_{3i}$ are closer in energy than some small $\epsilon$ of our choosing.  
For $\delta_s <0.17$, this can happen only if (a) $| g_{3i-m}-  g_{3i+n}| < \delta_w$ for $m,n=1,2$, or (b) if $\left| \sqrt{ g_{3i-2}^2 + g_{3i-1}^2 } \pm \sqrt{ g_{3i+2}^2 + g_{3i+1}^2 }  \pm g_{3i-2} \pm g_{3i+2}\right| < \epsilon$, which for a given choice of $g_{3i-2}, g_{3i+2}$ is no more probable than having $|g_{3i+1}  - g_{3i+1}| < \epsilon$.  
Thus resonances occur only if two of the couplings on nearby sites are nearly equal. 
Although the probability of single-site resonances is quantitatively different in our model as compared to Imbrie's, it is bounded by $\epsilon$.  Additionally, the probability of a resonant region of diameter at least $3L$ falls off exponentially in $L$.  This is the essential qualitative behaviour required to demonstrate MBL.  

(2) Structure of graph expansion: to demonstrate MBL, at each order in perturbation theory we must rotate away all couplings that arise from the $\tilde{X}_{3i}$ that are of the appropriate order in $\delta_w/\delta_s$.  Each rotation generates an infinite number of new couplings between sites that may be arbitrarily far apart (but whose strength falls off exponentially in the separation between sites).  The new couplings are described by a graph expansion, whose structure is fixed by the commutation relations between the operators $A_{3i}^{( \sigma, \sigma')} = g_{3i}\tilde{X}_{3i}^{( \sigma, \sigma')}/(E_\sigma - E_{\sigma'})$ where $\sigma, \sigma'$ denote the initial and final spin configurations.  In our case, $[ A_{3i}, A_{3j} ] =0$ if $|i-j| \geq 2$, exactly as in the Ising model in \cite{Imbrie:2016aa, Imbrie2016}; thus the structure of the graph expansion is the same (though the rank of the operator associated with each graph will be larger in our case, since the Hilbert space associated with each block of $3$ sites has dimension larger than $2$).  

These two similarities, together with the fact that the model satisfies limited level attraction, should be sufficient to ensure that the perturbation theory can be carried out to infinite order with the probability of resonant regions falling off exponentially in the size of these regions. Thus, we expect that the phase is MBL. 

We note that as for the MBL phase at large $W_z$, the $l$-bits in this case (given by the eigenvalues of $\{\hat{O}_{3i} \}$) are not algebraically independent, since only certain sequences of these may appear.  This is clear from the fact that though there are 5 possible eigenvalues for each $i$ (one of which is degenerate), the number of states in the Hilbert space grows only as $(\phi^3)^n \approx 4.24^n$ for $n$ clusters of $3$ spins.  We therefore anticipate that the structure of $l$-bits in this case is similar to that of the diagonal MBL phase discussed in the main text.  

\section{Supporting numerical data} 
\label{ThermApp}

Here we include further numerical evidence for thermalization, localization, and criticality in the regions of the phase diagram discussed in the main text.

First, we present a more detailed analysis of the diagonal MBL phase.  Here we expect the system to have $l$-bits -- conserved quantities that are related to on-site operators by a local unitary transformation.  
For large $W_z/W_x$, as discussed in the main text, we expect these $l$-bits to be well approximated by $\{ \tilde{Z}_i \}$ -- and indeed in this range $P(\langle \tilde{Z}_i  \rangle)$ is sharply peaked about $\pm 1$, as shown in the main text in Fig. 2(a).  
However, the minimum value of $P(\langle \tilde{Z}_i  \rangle)$ near $\langle \tilde{Z}_i  \rangle=0$ is only suppressed by two orders of magnitude as compared to the peaks at $\pm 1$, suggesting that even at $W_z/W_x =8$, $\{ \tilde{Z}_i \}$ are not close to the true $l$-bits for all $i$.
We conjecture that the probability weight near $\langle \tilde{Z}_i  \rangle=0$ comes from eigenstates where spins $i-1$ and $i+1$ point primarily along $+z$, so that spin $i$ points along/against its local field direction $\hat{u}_i$. 
As the local field is in the $x$-$z$ plane, this reduces the $z$-projection of spin $i$ in such eigenstates.

\begin{figure}[ht]
\centering
\includegraphics[width=\columnwidth]{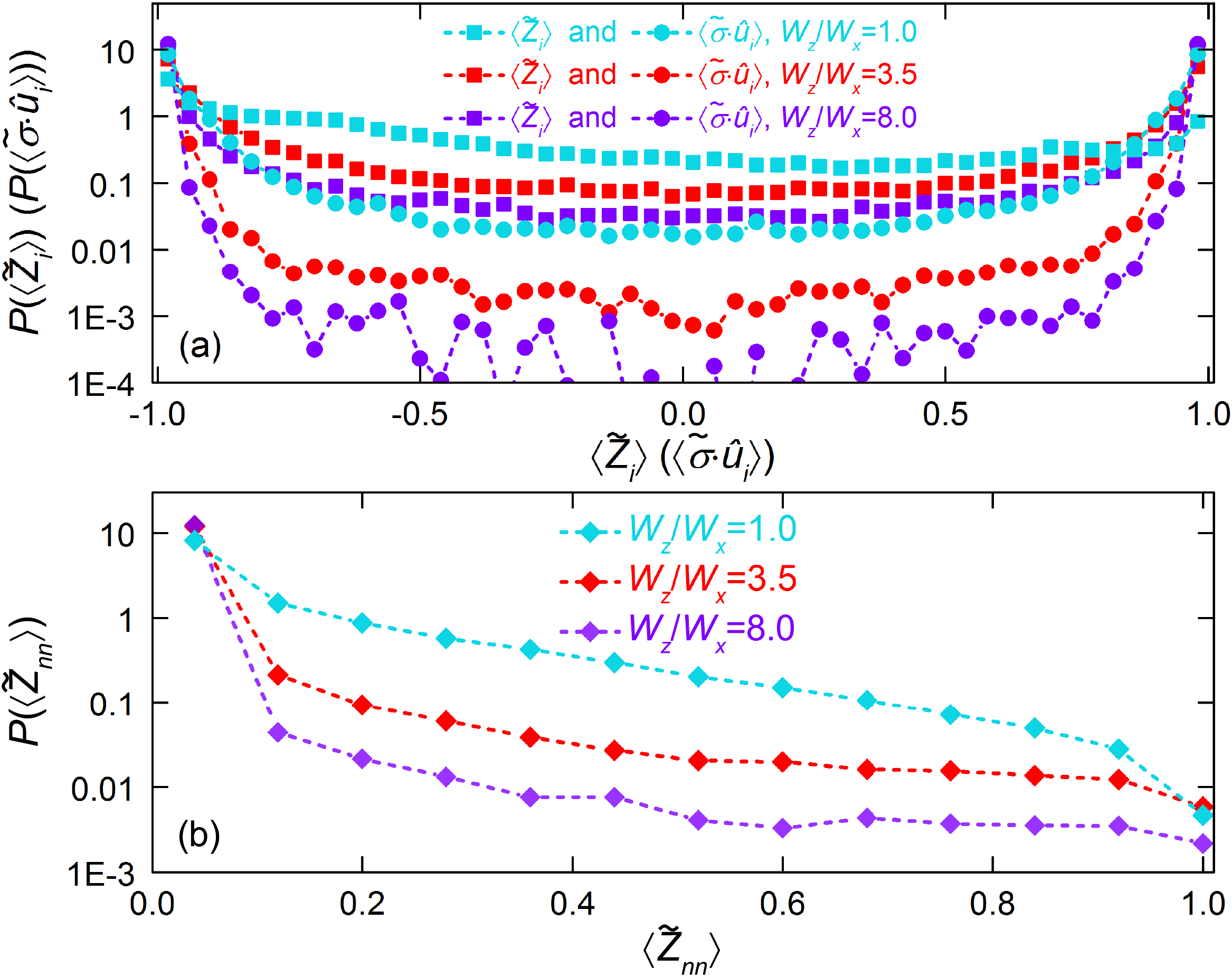}
\caption{\label{LbitsFigApp}  (a) Probability distributions of $\tilde{Z}_i$ and $\tilde{\sigma} \cdot \hat{u}_i$ at $N=20$ for different $W_z/W_x$. $P(\tilde{\sigma} \cdot \hat{u}_i)$ exhibits a stronger bi-modality as compared to $P(\tilde{Z}_i)$ showing that $\tilde{\sigma} \cdot \hat{u}_i$ is a better approximation to the $l$-bit operator $\tilde{\tau}_i^z$ as compared to $\tilde{Z}_i$. These properties persist to $W_z/W_x=1$ in the putative constrained MBL phase. (b) Probability distribution of $\langle \tilde{Z}_{nn} \rangle $ (see text for definition), showing that the 2-point function is very well approximated by a product of 1-point functions and a suppression of weight near $ \langle \tilde{Z}_{nn} \rangle=1$ due to the local constraint. }
\end{figure}

In order to test this conjecture, we compare the probability distribution of $\tilde{Z}_i$ with the probability distribution of the projection of spin $i$ onto the local field direction, $\tilde{\sigma} \cdot \hat{u}_i$ in Fig.~\ref{LbitsFigApp}(a) at $N=20$ and at different values of $W_z/W_x$. Deep in the diagonal MBL phase at $W_z/W_x=8$, $P(\tilde{\sigma} \cdot \hat{u}_i)$ exhibits a much stronger bi-modality as compared to $P(\tilde{Z}_i)$, confirming our conjecture that $\tilde{\sigma} \cdot \hat{u}_i$ is a better approximation to the $l$-bit operator $\tilde{\tau}_i^z$ as compared to $\tilde{Z}_i$. 
Decreasing $W_z/W_x$ takes the system into the putative constrained MBL regime; remarkably, $\tilde{\sigma} \cdot \hat{u}_i$ remains strongly bi-modal with weight primarily near $\pm 1$ and provides a good approximation to the $l$-bits even in this regime.
 
In addition, Fig.~\ref{LbitsFigApp}(b) shows the distribution of $\langle \tilde{Z}_{nn} \rangle$, defined as:
\begin{align}
\langle \tilde{Z}_{nn} \rangle \equiv \frac{(\langle E| \tilde{Z}_i |E \rangle  - 1) (\langle E| \tilde{Z}_{i+1} |E \rangle -1 )}{4}.
\end{align}  
Since the constraint implies the relation: 
\begin{align}
\langle E|( \tilde{Z}_i   - 1) ( \tilde{Z}_{i+1}  -1 )|E \rangle =0,
\end{align} 
we obtain a simple interpretation of $\langle \tilde{Z}_{nn} \rangle$ as the connected two point function:
\begin{align}
\langle \tilde{Z}_{nn} \rangle = \langle E| \tilde{Z}_i  \tilde{Z}_{i+1} |E \rangle - \langle E| \tilde{Z}_i |E  \rangle \langle E| \tilde{Z}_{i+1} |E \rangle.
\end{align}  
We expect the distribution of this quantity to be peaked near zero as the 2-point function is very well approximated by a product of 1-point functions in a localized phase.  
The local constraint further suppresses the weight near one, as:
\begin{align}
\langle  \tilde{Z}_{nn} \rangle = 1 \Rightarrow \langle E| \tilde{Z}_i |E \rangle = \langle E| \tilde{Z}_{i+1} |E \rangle = -1
\end{align}
which is forbidden by the local constraint ($\tilde{Z}_i$ and $\tilde{Z}_{i+1}$ cannot simultaneously equal $-1$).

\begin{figure}[ht]
\centering
\includegraphics[width=\columnwidth]{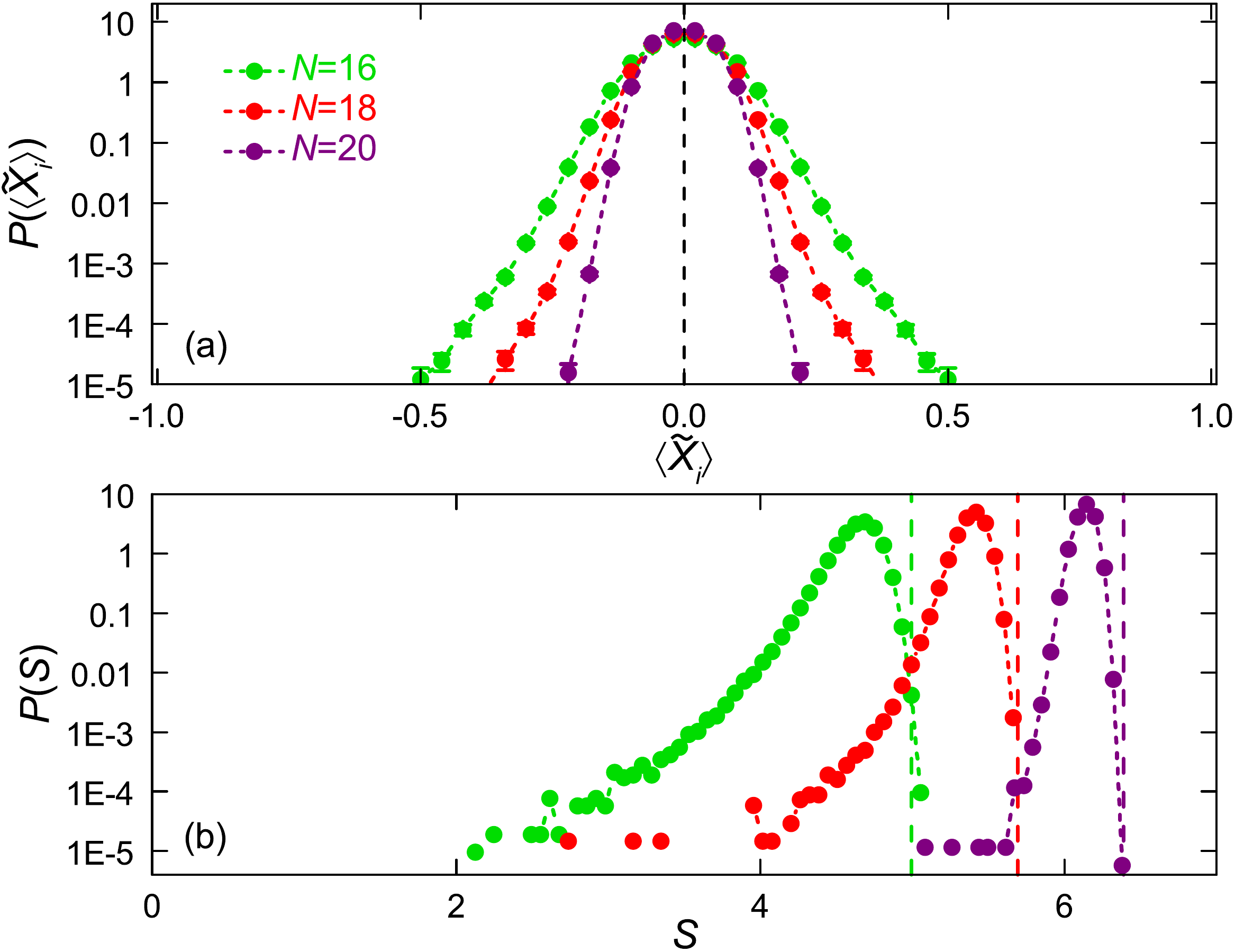}
\caption{\label{ThermFig}  Probability distributions of $\langle E| \tilde{X}_i |E \rangle$ (top), and the half-chain entanglement entropy $S$ (bottom)  at $g_x/W_x=3.5$, $W_z=0$ in the constrained thermal phase, with the vertical lines denoting their respective infinite temperature values of zero and $S_T$. The distributions narrow with increasing $N$ and the weight concentrates near the dashed lines.  }
\end{figure}

Next, we turn to the constrained thermal phase to test if eigenstate thermalization holds for generic observables.
In Fig. 2(c) in the main text, we showed that the distribution $P(\langle \tilde{Z}_i \rangle )$ narrows with increasing $N$ around the infinite temperature average $1/\sqrt{5}$.
In Fig.~\ref{ThermFig}, we test if $\tilde{X}_i$ and the half chain entanglement entropy $S$ also take their infinite temperature values when evaluated in eigenstates.
$P(\langle \tilde{X}_i \rangle )$ narrows around its infinite temperature average of zero as expected.
While $P(S)$ narrows with increasing $N$, the distribution is almost completely to the left of the infinite temperature value of $S_T \approx S_P - 0.06$ (see footnote [97] of the main text for the definition of $S_P$).
However, as the width of the distribution narrows with increasing $N$, we expect that the weight concentrates near $S_T$ as $N \to \infty$.

\begin{figure}[ht]
\centering
\includegraphics[width=\columnwidth]{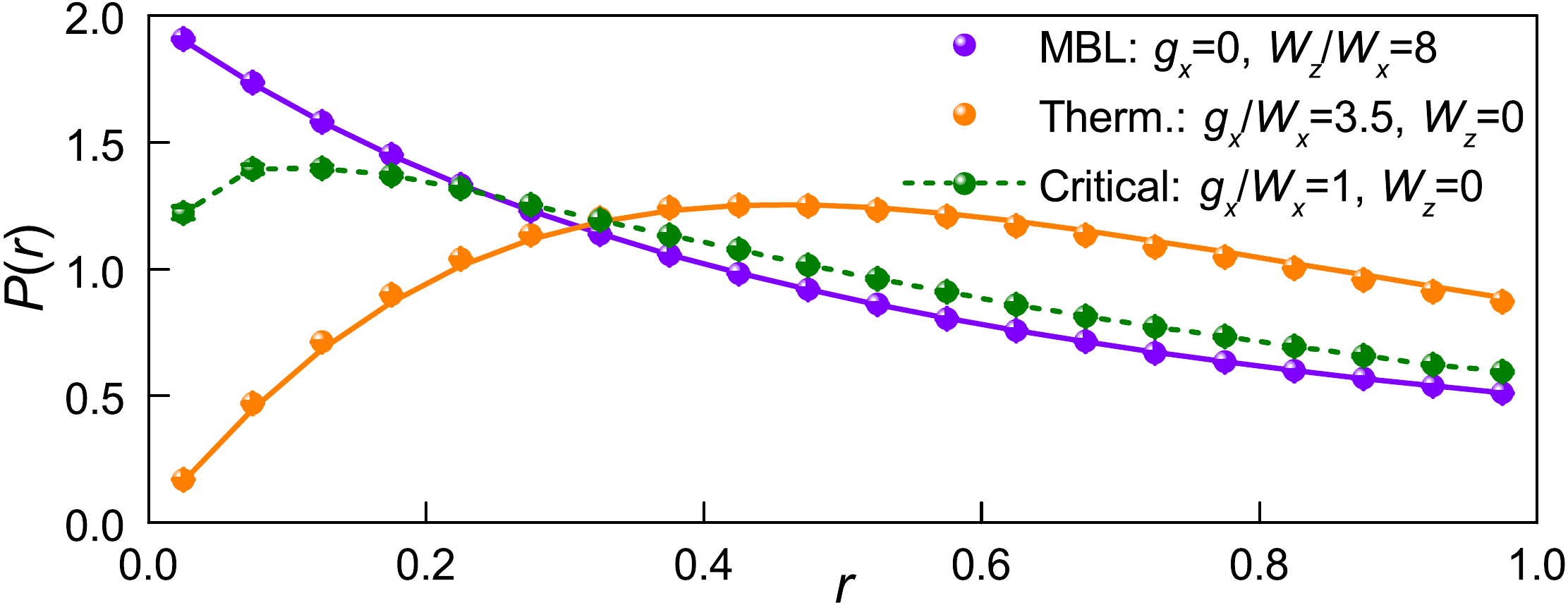}
\caption{\label{rAppFig}  Distribution of the level-spacing ratio $r$ for representative points in the constrained thermal (orange) and diagonal MBL (purple) phases, and at the dynamical transition (green) at $N=18$. Solid lines represent the corresponding distributions for the GOE (orange) and Poisson ensembles (purple). At the transition, $P(r)$ is suppressed at small $r$ similar to the GOE distribution, but the tail at $r \approx 1$ is close to the distribution of the Poisson ensemble.   }
\end{figure}

Finally, at the transition point at $g_x/W_x=1, W_z=0$, we find that most quantities (such as distribution of $r$, $S$, and $\langle \tilde{Z}_i \rangle$) behave in a manner that appears intermediate between thermalizing and localizing.
An example is shown in Fig.~\ref{rAppFig}, which compares the distribution of the the level spacing ratio $r$ at $N=18$ for representative points in the constrained thermal (orange) and diagonal MBL (purple) phases with that at the transition (green).  
In the thermalizing phase, the orange data points exhibit level repulsion ($P(r) \to 0$ at $r \to 0$) and lie on the solid orange line, which is the distribution expected for the random Gaussian Orthogonal Ensemble (GOE).
In the localized phase, the purple data points exhibit no level repulsion and lie on top of the solid purple line, which is the distribution expected for the Poisson ensemble with independent energies.
$P(r)$ at the transition is seen to be intermediate between the GOE and Poisson distributions: at small $r$, $P(r)$ is small indicating some level repulsion, while at large $r$, $P(r)$ approaches the curve corresponding to the Poisson ensemble. 
It is worth noting that the existence of the third distribution exactly at the metal-insulator transition was first proposed by Ref.~\cite{Shklovskii93} in the absence of interactions.

\section{Mapping to Dimer model}
\label{App:MapQDLConstrained}

\begin{figure}[h]
\centering
\includegraphics[width=0.85\columnwidth]{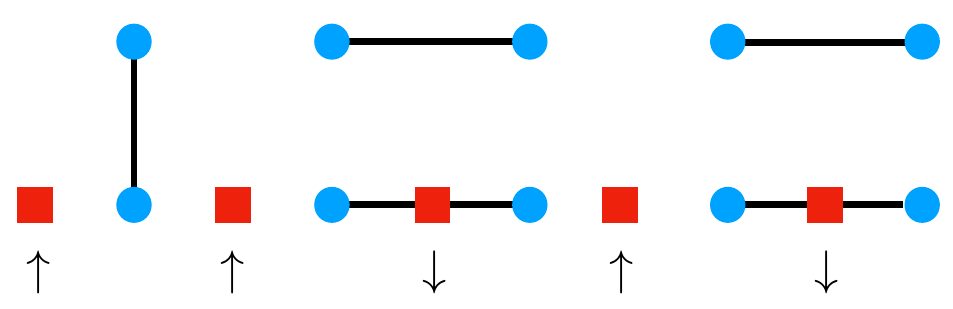}
\caption{\label{DimerFig} Ising configurations on the red spins map to dimer configurations on the ladder with blue vertices through the rules discussed in the text. The Ising spins satisfy the constraint that no two adjacent spins can point along $-z$, while the dimer configurations satisfy the constraint that there is exactly one dimer emanating from each blue vertex. }
\end{figure}

There is a one-to-one mapping between the constrained Ising configurations studied in this article and the configurations of a quantum dimer ladder.
This is how numerics has been done on dimer ladders \cite{Moessner:2001kl,Laumann:2012ab} and we present it here for completeness.

Dimer models naturally arise in low-energy theories of anti-ferromagnetic lattice spin systems \cite{Rokhsar:1988kx,Moessner:2001kl}.
Dimers reside on the bonds of the lattice and represent singlet states of the spins on the vertices that they connect.
As a given spin can at most participate in one singlet state, the dimer model Hilbert space comes with the natural constraint that each vertex has {\it{exactly one}} dimer (note that we excludes `monomers' or unpaired spins in this discussion).
A typical dimer model configuration on a two-leg ladder is shown in Fig.~\ref{DimerFig}, in which the dimers are represented by solid black lines and join the blue vertices.

The {\it{quantum}} dimer model makes the further simplification that distinct dimer model configurations are orthogonal to one another.
The ground state and low-energy properties of quantum dimer models have been extensively studied in bi-partite and non-bi-partite lattices since the late eighties, and they are known to exhibit a rich array of phases including crystalline ones and topological ones (see Ref.~\cite{Moessner:2008tg} for a recent review).

Each $z$-configuration of the constrained Ising chain is in one-to-one correspondence with a quantum dimer configuration on a two-leg ladder in the zero winding number sector.
Taking the Ising spins to reside on the midpoints of the bonds of the lower leg of the ladder, we place (1) two horizontal dimers on plaquette $i$ when $\tilde{Z}_i = -1$, and (2) one vertical dimer on the bond shared between plaquettes $i$ and $i+1$ when $Z_{i} = Z_{i+1}=1$. 
This produces the $z$-configuration for the red squares in Fig.~\ref{DimerFig}.
The dimer configurations have zero winding number as the number of horizontal dimers in any plaquette is even. Note that we have implicitly assumed open boundary conditions; with periodic boundary conditions, there is a second $\mathbb{Z}_2$ winding number associated with the parity of the total number of vertical bonds and the Ising configurations map to an appropriate sector of this winding number.

Finally, the operator $\tilde{X}_i$ maps to the usual resonance move on plaquette $i$:
\begin{align}
\tilde{X}_i \rightarrow |\, \|\,\rangle \langle =  | + | = \rangle \langle \, \| \, | 
\end{align}
while the operator $\tilde{Z}_i$ maps to the potential cost:
\begin{align}
\tilde{Z}_i \rightarrow 1 - 2 | = \rangle \langle = |
\end{align}
Thus, the results in this article apply \emph{mutatis-mutandis} to the dynamics of a class of quantum dimer ladders.

\end{document}